\documentclass[runningheads]{llncs}
\usepackage{graphicx}
\usepackage[table,xcdraw]{xcolor}
\usepackage{multirow}
\usepackage{subfig}

\begin{document}
\title{Comparison of Online Maneuvers by Authentic and Inauthentic Local News Organizations\thanks{This work was supported in part by the Office of Naval Research (ONR) Award N00014182106, the Knight Foundation, the Center for Computational Analysis of Social and Organizational Systems (CASOS), and the Center for Informed Democracy and Social-cybersecurity (IDeaS). The views and conclusions contained in this document are those of the authors and should not be interpreted as representing the official policies, either expressed or implied, of the ONR or the U.S. government.}}
\titlerunning{BEND Maneuvers by Local News and Pink Slime}
%
\author{ Christine Sowa Lepird\inst{1}\orcidID{0000-0002-9350-5627} \and
Dr. Kathleen M. Carley\inst{1}\orcidID{0000-0002-6356-0238}}
\authorrunning{C. Lepird et al.}
%
\institute{Carnegie Mellon University, Pittsburgh PA 15213, USA 
\email{csowa@andrew.cmu.com}\\
\url{http://www.casos.cs.cmu.edu} }
\maketitle              
\begin{abstract}

 Inauthentic local news organizations, otherwise known as pink slime, have become a serious problem exploiting the trust of local news since their creation ahead of the 2020 U.S. Presidential election. In this paper, we apply the BEND framework, a methodology of classifying social media posts as belonging to sixteen network and narrative maneuvers, to compare and contrast how pink slime sites and authentic local news sites are shared on Facebook Pages. It finds that pink slime sites implemented more positive narrative maneuvers than those of local news sharers. Both news types utilized distraction but to fulfill separate goals - pink slime used it against local and state elections while authentic local news focused on national elections and figureheads. Furthermore, local news employed the neutralize tactic in order to reduce positive sentiment around national politicians. 

\end{abstract}
\section{Introduction}

Leading up to the 2020 U.S. Presidential Election, a number of seemingly hyper-local websites emerged; however, these sites had no reporters in the local community. Additionally, many of these sites were owned by single organizations who were interested in spreading the same (often national) messaging to local communities, exploiting their trust in local reporting.  The journalist Ryan Smith coined the phrase ``Pink Slime" to describe these sites that are filled with low-quality, largely automated reporting and partisan political messaging that are trying to pass as local news and influence voters in elections\cite{tarkov_journatic_2012}.

There are over 1,000 of these pink slime sites, and they represent a problem due to the high level of trust Americans report in what they perceive as local news. Americans trust in national media has declined in the past 6 years, but Americans of all political leanings have retained a high level of trust in local news reporting \cite{gottfried_partisan_2021}.   Furthermore, while 20\% of local newspapers have closed their doors in recent years \cite{takenaga_more_2019}, the owners of these inauthentic local sites are filling the space with information that is not original nor human-produced.  

Existing literature largely analyzes the pink slime websites in the context of the 2020 U.S. Presidential election, through scraping articles on the websites for topic modeling \cite{royal_local_2017} or analyzing the web traffic of a sample of Americans in the run-up to the election \cite{moore_consumption_2023}.  One of the takeaways from the consumption studies was that Facebook is the largest social media referrer of visits to pink slime sites, accounting for 23.4\% of visits to these websites in 2020 \cite{moore_consumption_2023}; however, research into the tactics employed by these sites to garner these referrals on Facebook are lacking. Only one paper looking into referrals to pink slime sites from social media has been published \cite{burton_research_2020}; it analyzed data from Reddit and 4chan and concluded that pink slime sites are not shared on these platforms. However, the Reddit analysis was limited to data collected from top political subreddits, not subreddits targeting regional communities. Furthermore, it failed to include any Facebook data in the analysis. 

At the same time as the creation of pink slime sites, there has been increased scrutiny of the ownership of news media. Larger companies have purchased multiple local news outlets \cite{noauthor_who_nodate}, and casual observers of the pink slime phenomenon may be quick to group the issues together. This paper will illustrate the differences between local news owned by a larger parent company with authentic local reporters and pink slime journalism. It also analyzes the methods and maneuvers the organizations controlling the pink slime sites are employing to influence local communities whose votes are important to local, state, and national elections. It applies the BEND Framework to compare the tactics of these sites to those of more credible local news sites.

\subsection{Background and Related Work}

Priyanjana Bengani, a senior research fellow at Columbia Journalism School's Tow Center for Digital Journalism published the most prominent and comprehensive analysis of the origins of the pink slime sites and how they are controlled by a few organizations \cite{bengani_hundreds_2019}.  The largest of these organizations, Media Metric, controls over 900 sites targeting local areas in the country. 

Bengani later collaborated with researchers at Stanford to see how these sites were consumed during an election. They found that leading up to the 2020 U.S. Presidential Election 3.7\% of Americans were exposed to news from these sites \cite{moore_consumption_2023}, with Facebook being one of the primary referrers to the sites. While this figure is lower than the 39.1\% and 36.4\% who were exposed to misinformation and local news sites, respectively, during the same time period, the number of pink slime sites and organizations have grown. While 22\% of the pink slime articles at the time had to do with local gas prices, very few individuals visited these sites; rather, the most consumed pink slime news articles pertained to politics, despite there being very few articles on the topic \cite{moore_consumption_2023}. Furthermore, they found that while supporters of Donald Trump were significantly more likely to visit misinformation sites, Biden supporters were significantly more likely to visit pink slime sites.

Researchers at Duke University analyzed the articles on the Metric Media websites for 78 days in 2020 to categorize and analyze the contents of what they were promoting \cite{royal_local_2017}. Using a Jaccardian similarity index, the researchers analyzed the text content of stories appearing on the multiple sites and found that, on average, 92\% of the text of an article on one Metric Media site appeared in the same story on another site - largely swapping a few words \cite{royal_local_2017}. When analyzing the authorship of the non-autogenerated stories on Metric Media sites, they found that 
46\% of Metric Media websites contained no articles written by a human; the remaining 54\% of Metric Media-owned sites had stories that were attributed to 161 human reporters, with each reporter averaging 26 articles across a median of 9.5 states each \cite{royal_local_2017}. The ten most published reporters averaged 236.5 articles, and the median age of a front page story was 81 days \cite{royal_local_2017}. However, this research (like others mentioned earlier) focuses only on data from a three month period in 2020. 

With this research in mind, we incorporate pink slime articles shared to Facebook as recent as May of 2023 and compare the narrative and network maneuvers they are utilizing in comparison to those of authentic local news sites to understand the differences in how they are being promoted on social media.

\section{Data and Methods}

\subsection{Data Collection and Cleaning}

\begin{figure}%
    \centering
    \subfloat[\centering Local News]{{\includegraphics[width=5cm]{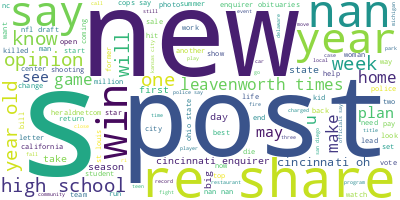} }}
    \qquad
    \subfloat[\centering Pink Slime]{{\includegraphics[width=5cm]{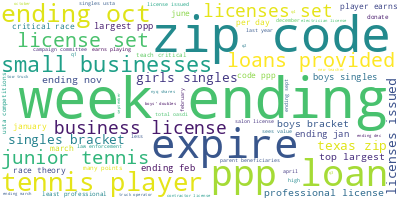} }}%
    \caption{Wordcloud of all news headlines}%
    \label{fig:wordclouds_all}%
\end{figure}

To find the news site domains we were interested in studying, the researchers consolidated a list of known  pink slime sites \cite{tow_center_for_digital_journalism_domains_nodate} as well as the list of authentic local news sites owned by larger companies \cite{noauthor_who_nodate} - this included Black Press Group, Digital First Media, Gannett, Lee Enterprises, New Media, Raycom Media Inc., The McClatchy Company, and tronc Inc. Using the CrowdTangle API \cite{noauthor_crowdtangle_2020}, for each of the domains on the list, the 1,000 most recent instances of a link to the domain being shared on a Facebook Page was collected. In total 335,609 posts were collected from 12 pink slime organizations and 8 local news organizations. Of the 12 pink slime organizations, there were 1,238 domains linked to from 285,640 posts. Of the 8 local news organizations, there were 50 domains linked to by 49,969 posts. Using the wordcloud package in Python \footnote{pypi.org/project/wordcloud/} to analyze the difference in the words used in the headlines for local news and pink slime (illustrated in Figure \ref{fig:wordclouds_all}), we see the words most used in local news are either 

\begin{figure}%
    \centering
    \subfloat[\centering Local News]{{\includegraphics[width=5cm]{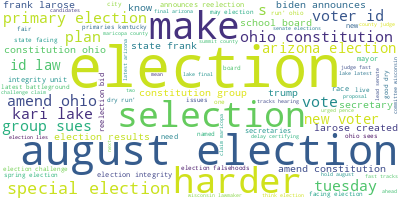} }}
    \qquad
    \subfloat[\centering Pink Slime]{{\includegraphics[width=5cm]{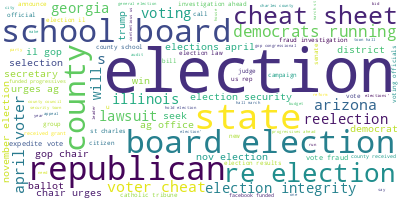} }}%
    \caption{Wordcloud of news headlines pertaining to elections}%
    \label{fig:wordclouds_elections}%
\end{figure}

After performing topic modeling on the titles of the shared links, the largest common topic found pertained to elections. Since research shows that the most consumed pink slime sites are those pertaining to politics \cite{moore_consumption_2023} and in order to analyze how these two groups discussed the same topic, the posts were filtered down to ones mentioning elections, judicial selections, and voting. This left 385 posts linking to 47 local news domains and 465 posts linking to 76 different pink slime domains. The local news posts ranged from November 17, 2022 to January 12, 2023. The pink slime posts ranged from January 27, 2020 to May 12, 2023; unsurprisingly, the date range is larger due to the lack of timely reporting \cite{royal_local_2017} by these sites. The posts linking to local news sites averaged a higher number of likes (average 27.2, standard deviation 124.7) than that of pink slime (average 7.3, standard deviation 59.4), which follows the previous research which indicates local news sites receive higher levels of consumption than pink slime sites \cite{moore_consumption_2023}. Wordclouds generated from the elections-specific dataset (as seen in in Figure \ref{fig:wordclouds_elections}), show some similar words appearing on the same topic, discussing dates (``august", ``tuesday" ``april") and states where the elections are taking place; however, pink slime headlines have a higher prevalence of phrases such as ``cheat sheet", ``republican", and ``democrats".  While this information shows some differences and similarities in the words in the headlines, it fails to illustrate \textit{how} these words are being used (the insights we will find by applying the BEND framework).

\subsection{Linguistic Cues}

Once the Facebook post data was collected, the text of the posts was analyzed for specific linguistic features, referred to as ``cues", using the NetMapper software \cite{carley_ora_2018}. NetMapper analyzes each segment of text in a post (in this case, the `title' attribute from CrowdTangle which auto-populates as the headline of the linking news article) to estimate various emotional states. The software appends metadata about the post's sentiment based on the number of terms specific to certain emotions, use of pronouns, and icons present. Ultimately, 121 cues are appended to each post based on this in-depth analysis. The cues extracted from the post help to classify which BEND maneuvers are taking place.

\subsection{Organization Risk Analyzer - Pro Software}

The Organization Risk Analyzer (ORA)-PRO software \cite{carley_ora_2018} was used to visualize and analyze the dynamic meta-networks present in this dataset. The local news and pink slime datasets were loaded into ORA as separate meta networks. 

\begin{figure}%
    \centering
    \subfloat[\centering Local News]{{\includegraphics[width=5cm]{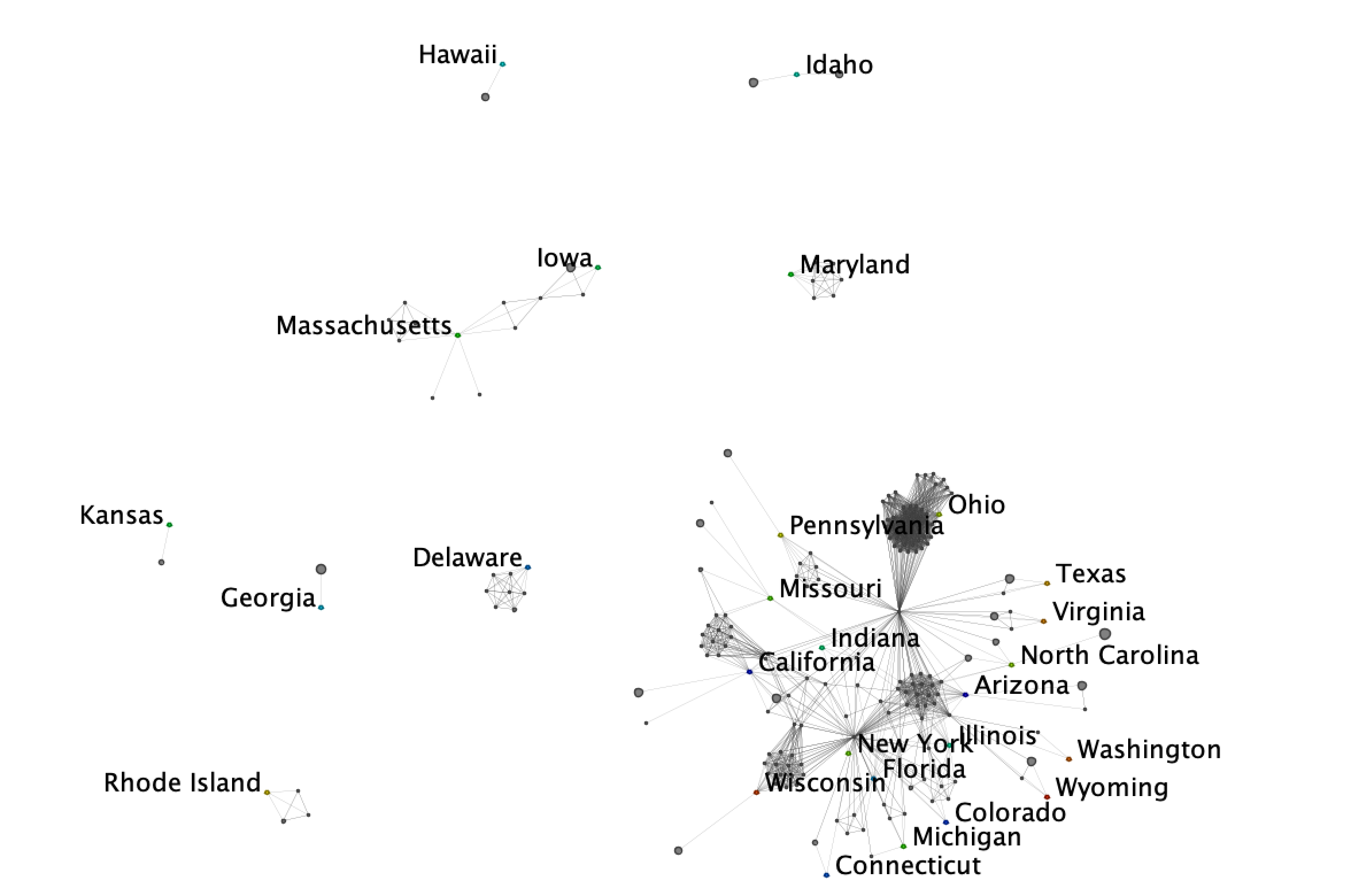} }}
    \qquad
    \subfloat[\centering Pink Slime]{{\includegraphics[width=5cm]{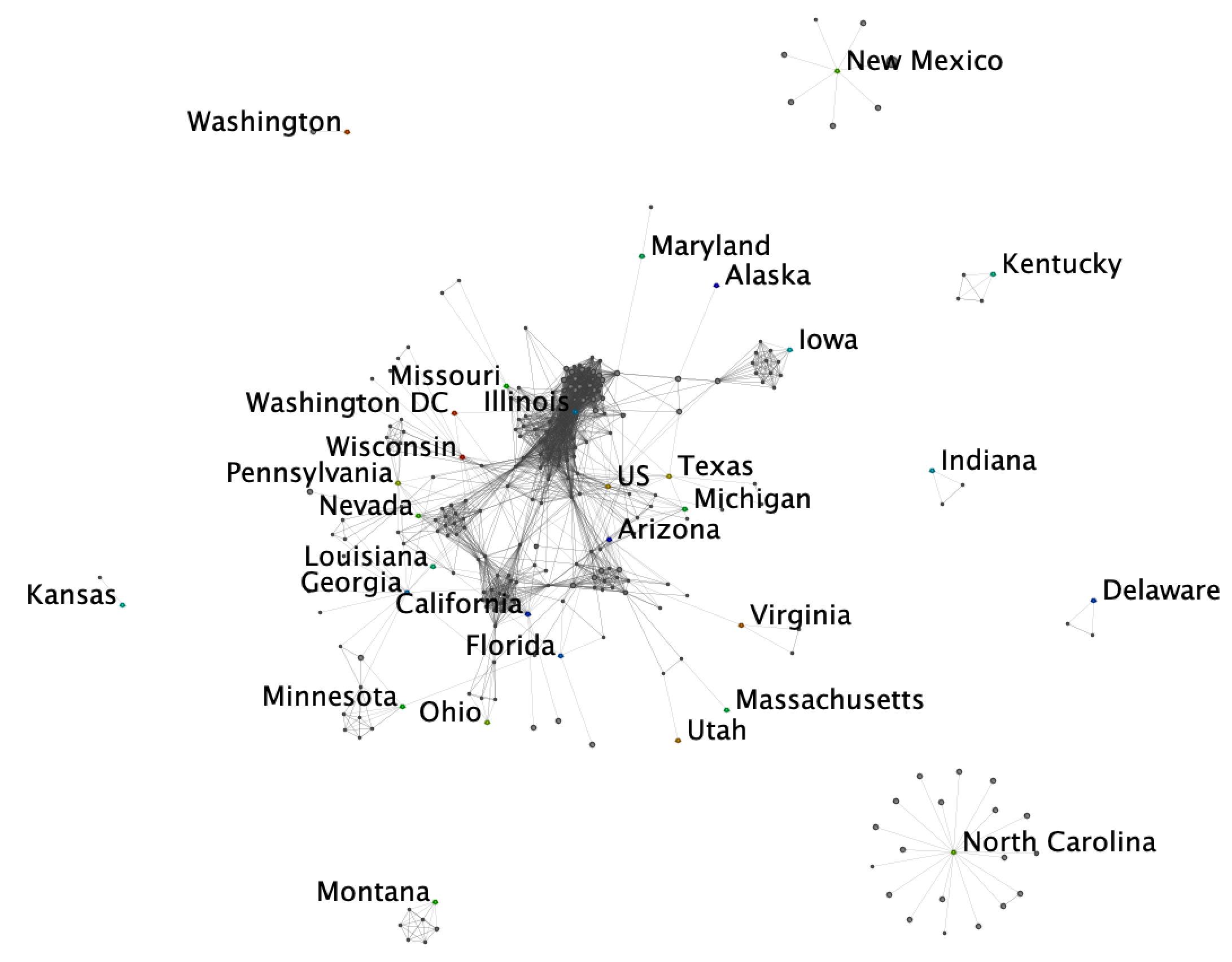} }}%
    \caption{Network visualization of Facebook Pages x Shared Website Location (State)}%
    \label{fig:networks}%
\end{figure}

Visualizing the networks of Facebook Pages by the state their local news or pink slime site shared corresponds to (as seen in Figure \ref{fig:networks}) can glean some insight into the network operations of the two news types. 

While authentic local news, Ohio has the most Facebook Pages (145) linking to news sites.  For the network of Facebook Pages sharing pink slime, we see that Illinois has the highest total degree centrality, with over 300 Facebook Pages linking to Illinois pink slime sites. This isn't surprising given there is a sub-organization of the Metric Media organization dedicated to creating pink slime sites called LGIS (this arrangement exists for no other states).  For some states like Kansas and Delaware, regardless of whether Facebook Pages are sharing pink slime or local news, those same pages didn't sharing local news or pink slime sites of others' states. For this network, we find that local news has a higher density (0.15) than that of pink slime (0.12). However, much like our analysis of the word clouds, these networks do not provide the needed insight into the full picture - how did the networks facilitate the conversation taking place within the posts? 

In order to answer that question, we use ORA to run the BEND and Community Assessment report on the Facebook posts and their linguistic cues. The network used to determine the network maneuvers of these posts are the Facebook Pages x News Domain x Facebook Pages networks, indicating that there is an ``interaction" if two Facebook Pages share posts linking to the same news site; the density of this interaction network was 0.02.

\subsection{The BEND Framework}

\begin{figure}
\begin{center}

  \noindent
  \includegraphics[scale=0.3]{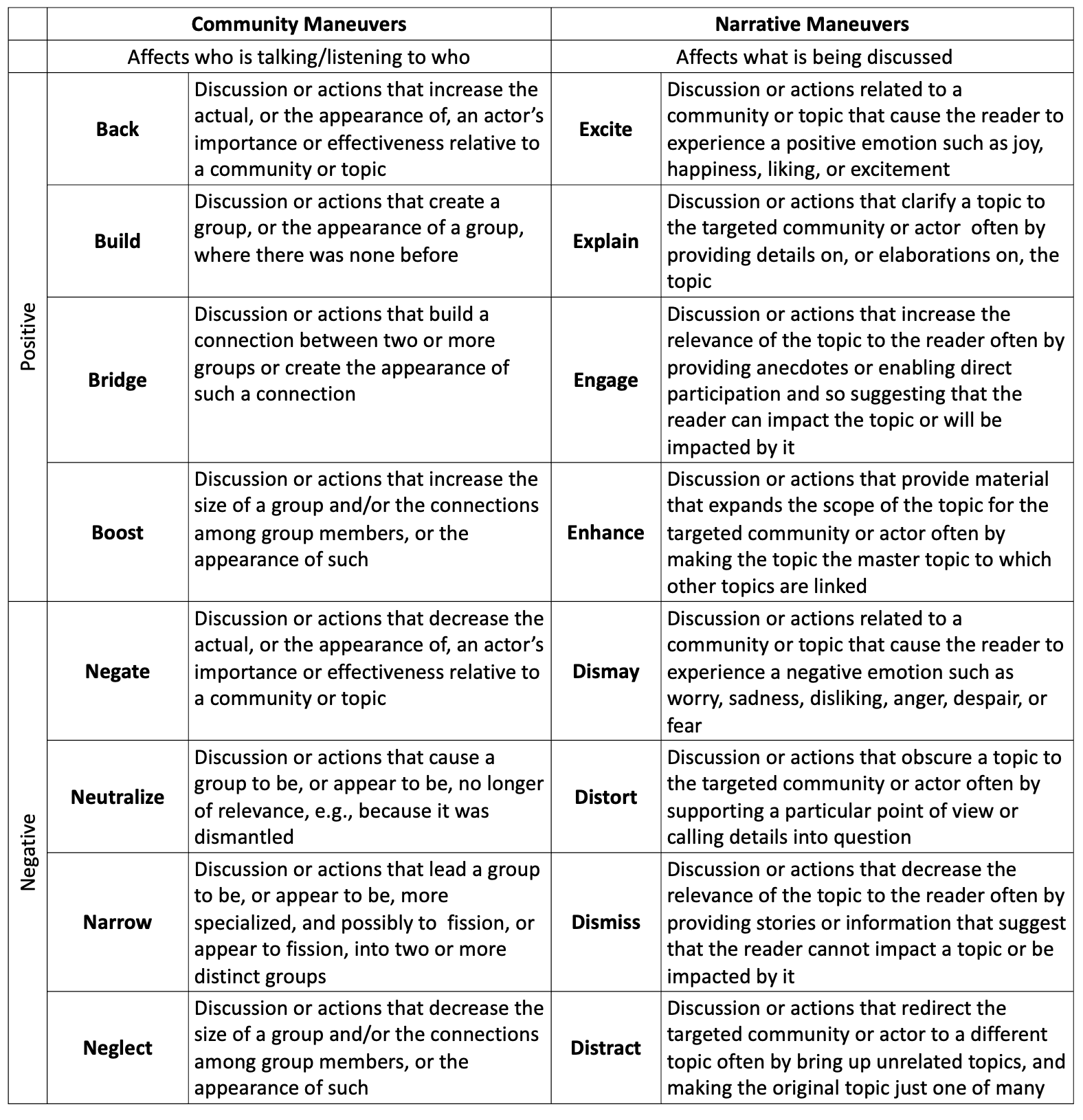}
  \caption{Definitions of the 16 BEND Maneuvers, adapted from \cite{armySocialCybersecurity}, \cite{blane_social-cyber_2023}, and discussions with the authors.}
  \label{fig:bend-defs}
\end{center}
\end{figure}

The CASOS Center at Carnegie Mellon University has produced substantial research in the field of categorizing online influence operations; they have published a set of 16 defined maneuvers utilized in influence operations, referred to as the BEND framework \cite{carley_bend_2020}.   The 16 categories can be broken into narrative (based on the text messaging and the way in which it is presented) and network (based on the way in which the messaging is spread and communities are formed around the key actors) maneuvers. Each of the letters includes four maneuvers of the same starting initial. The B maneuvers (Back, Build, Bridge, and Boost)  are positive network maneuvers. The E maneuvers (Engage, Explain, Excite, and Enhance) represent positive narrative maneuvers. The N maneuvers (Neutralize, Negate, Narrow, and Neglect) are utilized via negative network means. Finally, the D maneuvers (Dismiss, Distort, Dismay, and Distract) are negative narrative maneuvers. Their individual definitions can be found in Figure \ref{fig:bend-defs}. This framework allows for a more defined, measured, and analytical way to compare ways in which influence tactics are employed in information operations.

While BEND has largely been utilized for analyzing behavior on Twitter (such as narratives around pro and anti-vaccination campaigns \cite{blane_social-cyber_2022} \cite{blane2023analyzing}, Indonesian information operations campaigns \cite{danaditya2022curious}, and elections), this research will implement the methodology to categorize the maneuvers of sharers of pink slime and local news through Facebook. 

When BEND is applied to Twitter data, the networks that the maneuvers were built on were for User x User by shared hashtag, retweet, or reply. Due to the way in which Meta shares Facebook data via CrowdTangle, information about direct relationships between Facebook Pages was unavailable. Instead, the network that was used for this study (Facebook Page x Domain x Facebook Page) is more limited because it does not imply a direct interaction between the two users.

\section{Analysis and Results}
\subsection{Data Analysis}

\begin{figure}
  \noindent
  \includegraphics[width=\textwidth]{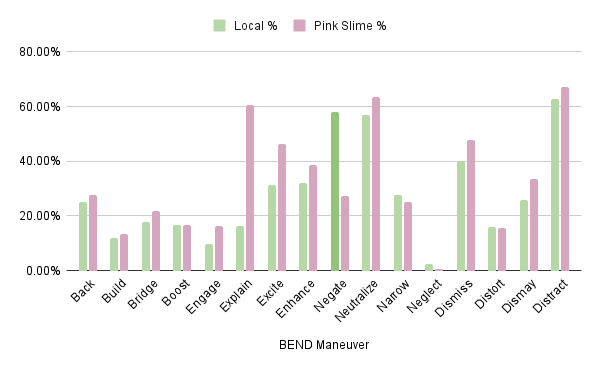}
  \caption{Percentage of Posts Using BEND Maneuvers by News Type}
  \label{fig:bend-narrative}
\end{figure}

Table \ref{fig:bend-narrative} illustrates the percentage of Facebook posts that contain each of the BEND Maneuvers (a note that a post can contain multiple BEND Maneuvers). 

Both groups had over half of their messages falling in the Distract category. While less than 20\% of documents had each of the B maneuvers, the percentages utilized by local news and pink slime are fairly equal. 

\begin{figure}
  \noindent
  \includegraphics[width=\textwidth]{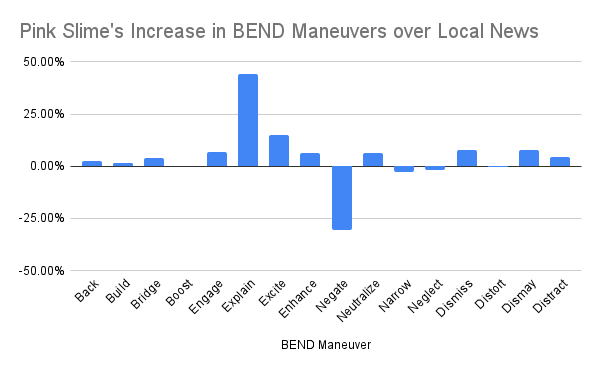}
  \caption{Increase in pink slime posts using BEND maneuvers over local news posts}
  \label{fig:bend-narrative-diff}
\end{figure}

Table \ref{fig:bend-narrative-diff} takes the values from Table \ref{fig:bend-narrative} and subtracts the local news values from the pink slime values. This shows how much more the pink slime posts are utilizing each BEND maneuver more than the local news posts.  

Most interestingly, many more pink slime posts utilize the Explain, Excite, Neutralize, and Dismiss maneuvers than local news. Local news posts, however, were more heavily involved in the Neutralize maneuver. 

For those sharing pink slime sites, the \textbf{Explain} maneuver can be seen in titles like:
\begin{itemize}
    \item ``Ninety-three percent of Arizona Catholics say religion should not play a factor in judicial selection"
    \item `Townsend: Audit of secretary of state’s use of private funds in elections necessary `to feel good about yes vote' on budget" 
    \item ``Grand Canyon Times requests communications between Gov. Hobbs and outside legal group involved in elections task force."
    \item ``Governor Walz Congratulates Departing Commissioner Roberts-Davis, Opens Selection Process for Cabinet Vacancy"
\end{itemize}  

The text of these posts convey statistics or quotes that provide insight into the topic. 

Meanwhile the messaging around \textbf{Excite} can be seen in posts like:

\begin{itemize}
    \item ``Allen: `We must restore our trust in the election process'" 
    \item  ``Coyne: `We are thrilled with this year’s local election results and are very proud of whatever impact we had in producing them’"
\end{itemize}

Much like with the Explain posts, the titles for Excite rely heavily on quotes. The narrative is one meant to bring positive emotion towards the audience. More than half of the messages fall into the Explain and Excite categories, keeping a majority of the messaging \textit{positive} in sentiment. The remainder of maneuvers analyzed fall into the categorization of \textit{negative} in their influence.

Examples of pink slime sites being shared with a \textbf{Neutralize} message include:
\begin{itemize}
    \item ``Arizona legislators protest election results, request decertification"
    \item ``Kansas legislature overrides Kelly's veto of election integrity bill."
\end{itemize}

When the \textbf{Dismiss} maneuver is analyzed for the pink slime sites, examples include:
\begin{itemize}
    \item ``Harbin: Georgia is experiencing `more election irregularities because our Secretary of State could not get the job done'"
    \item ``Nagel: `Democrats in Springfield are offering temporary election year gimmicks that attempt to trick voters instead of truly help them'"
\end{itemize} 

The last of which links to an article owned by the LGIS pink slime organization targeting a small city in Illinois. By referring to the state's capitol (Springfield), it gives the appearance of local news coverage; however, the same author also wrote articles for a different pink slime organization, Media Metric, targeting Grand Haven, Michigan. These Dismiss campaigns are aimed at minimizing the efforts of individuals or groups.

When the maneuvers for local news are analyzed, \textbf{Negate} (the largest increase over pink slime) is seen in messages like:

\begin{itemize}
    \item ``Trump: People who think 2020 election was fair are `very stupid’"
    \item ``Donald Trump’s response to criminal charges revives election lies" 
    \item ``School elections are now political: NYC Community and Education Council voting is getting too nasty."
\end{itemize}

Broadly, these messages are designed to reduce positive messaging on a topic or individual.

Both of the groups utilized the \textbf{Distract} maneuver heavily, a narrative maneuver that attempts to make other topics seem more important through misdirection. For pink slime this was seen in messaging like:
\begin{itemize}
    \item ``Rats and needles hot election issue in Rogers Park Aldermanic race" 
    \item ``Kansas challenger for secretary of state: Opponent's refusal to sign election integrity pledge `should be a red flag for any Republican voter'"
\end{itemize}

In local news, \textbf{Distract} looks like:
\begin{itemize}
    \item ``Biden launches 2024 campaign; jury selection to start in Trump rape lawsuit; N. Dakota's near total abortion ban; and more morning headlines" (which linked to an Idaho-based local news site)
    \item ``Did they vote twice in the 2022 election? RI investigating 5 cases of potential double voting."
\end{itemize}

\section{Discussion}

For both sites controlled by pink slime organizations and sites controlled by organizations owning multiple local news domains, the top-ranking BEND maneuver implemented was Distract - a negative narrative maneuver. However, pink slime sites used distraction in messaging pertaining to local and state elections while the local news sites had a greater focus on national elections and events in other regions. Perhaps authentic local news sites have enough established trust with their audience that they do not worry that viewers of their headlines will be concerned with reporting about what is happening in other regions. Furthermore, it is exemplified in their headlines that are daily `summaries', summarizing multiple (unrelated) events in a day into a single title. 

\begin{figure}
\begin{center}
  \noindent
  \includegraphics[width=8cm]{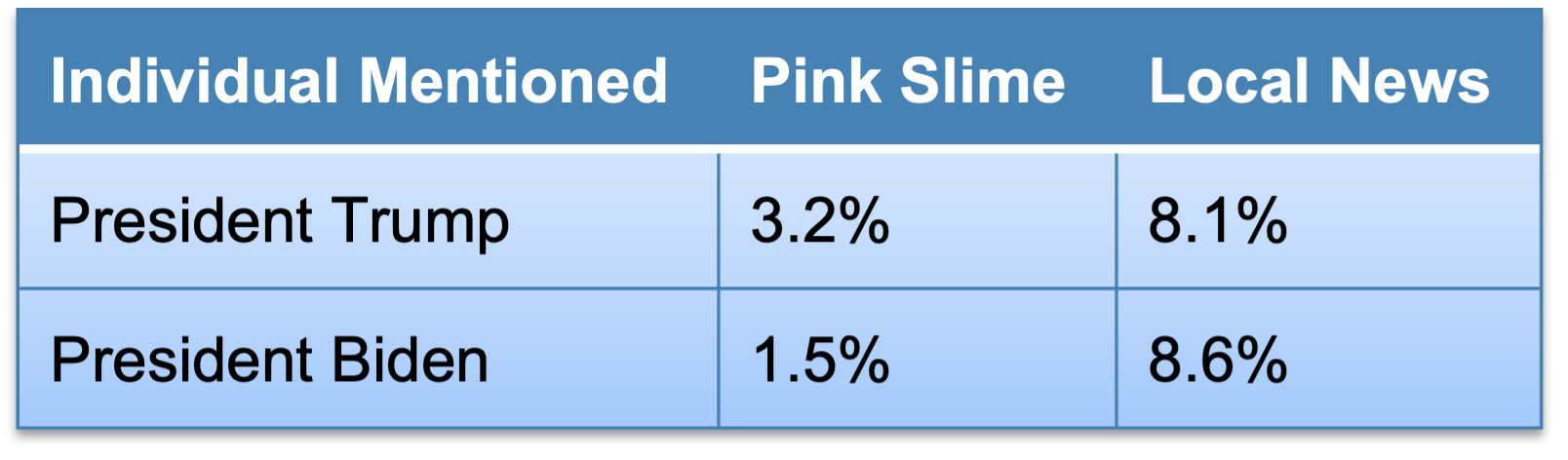}
  \caption{Percentage of article headlines referencing the current and past president by news type}
  \label{fig:presidents}
\end{center}
\end{figure}

Surprisingly, mentions of former President Trump were see in 3.2\% of posts linking to pink slime sites, but he appeared in 8.1\% of local news headlines; current President Biden was mentioned in only 1.5\% of pink slime sites but in 8.6\% of pink slime text, as illustrated in Figure \ref{fig:presidents}. This may indicate a concerted effort on the part of the pink slime sites to appear more local and less concerned with national politics. On the flip side, authentic local news reporters may know that their community is interested in how the actions of national figures affect their local community. 

Interestingly, sites controlled by pink slime organizations were shared on Facebook with more positive messaging than posts from local news organizations. Explain and Excite were utilized to highlight facts and nuance from both hyper-local and national political topics. This may be a consequence of creating API-driven automated news that pulls statistics and quotes from databases; it creates the illusion that these articles have explanatory insights and are politically neutral in a data-driven fashion.

When pink slime sites used negative messaging through Dismiss, out-of-state reporters highlighted reasons of local concern to dismiss efforts by political figures. Furthermore, these headlines use quotes from local politicians, potentially to show that the news source is invested in the local community (despite not having local reporters stationed there). 

Facebook Pages sharing local news sites heavily deployed the Negate maneuver to limit the appearance of the importance of national politicians and local organizations. It would appear that local news organizations aren't afraid to use this negative network maneuver to tarnish the image of controversial figures. 

\section{Limitations and Future Work}

Previous BEND publications focus on Twitter data and utilize the retweet network as a way of building the network ties. When applying the BEND framework to Facebook data, there are substantial limitations in the network elements due to the available data features from CrowdTangle. Individual Facebook Page account interactions with other Facebook Pages - re-posts, mentions, etc - would make for a more similar comparison to the Twitter network but are not included in the API. This work considered interaction to be when two Facebook Pages posted news articles to the same domains which does not indicate direct knowledge of other accounts but at least offers a proxy for shared knowledge. 

Another limitation is that these posts are not all referring to the same elections, and it is possible that some of the news domains have shifted their strategies over time. Future work will address this issue by focusing on a specific election, the 2020 U.S. Midterm Election. 

Finally, the pink slime sites studied in this article are all affiliated with the Metric Media organization. Since the original data collection, more pink slime news networks have been uncovered with differing political leanings. In the Midterms analysis, we will compare the BEND maneuvers utilized by the differing organizations to see if they have more in common or are implementing different tactics to influence public opinion. 

In the future study of Midterms posts, we will analyze similarities and differences in network and narrative maneuvers utilized by accounts sharing pink slime sites, local news sites, verified real news sites, and low credibility news sites across three social media platforms - Facebook, Twitter, and Reddit. This should provide a more substantial comparison to understanding the effectiveness of the Facebook network in comparison to Twitter's retweet one.

\section{Conclusions}

This research illustrated that the BEND Framework can be applied to data from the Facebook platform. Furthermore, it found similarities and differences in the way in which Facebook Pages utilized the BEND maneuvers to share pink slime sites and local news sites. While both pink slime and local news most heavily deployed the Distract maneuver, they did so with separate objectives. Local news will use it by lumping unrelated stories together in a headline discussing national and state politics. However, for pink slime, the primary topics are of hyper-local importance. 

Pink slime sites also had high incidence of Explain, Excite, and Dismiss while local news relied heavily on Negate (largely to reduce positive messaging around national politicians). While the pink slime sites highlighted the local cities and states that the sites target in their messaging, the sites owned by authentic local news organizations had a greater element of national politics and focus on other states.

%
%
%
%

\bibliographystyle{splncs04}

\bibliography{pink_slime_bend}

\end{document}